\documentclass{jfm}
\usepackage[utf8]{inputenc}
\usepackage{graphicx}
\usepackage{epstopdf, epsfig}
\usepackage{amsmath,amssymb,amsfonts,amssymb}
\usepackage{subfig}
\usepackage{setspace}
\usepackage{float}
\usepackage{microtype}
\usepackage[normalem]{ulem}
\usepackage{ragged2e}
\usepackage{xcolor}
\usepackage{tabularx}

\DeclareCaptionJustification{justified}{\justifying}
\usepackage{hyperref}
\hypersetup{
	hyperindex,
	breaklinks,
	colorlinks=true,
	linkcolor=blue,
	citecolor=magenta,
	bookmarks=true,
	bookmarksopen=true,
	bookmarksopenlevel=2,
	pdfstartpage={1},
	pdfstartview={FitH},
	pdfview={FitH 0},
	pdfauthor={B. F. Farrell, P. J. Ioannou, and M.-A. Nikolaidis},
	pdftitle={}}

\usepackage{ifthen}

\def\U{\bm{\mathsf{U}}}
\def\Uv{\mathbf{U}}

\def\P{{\bf P}}

\def\U{\bm{\mathsf{U}}}

\def\Uv{\boldsymbol{U}}

\def\P{{\bf P}}

\newcommand\redsout{\bgroup\markoverwith{\textcolor{red}{\rule[0.5ex]{2pt}{0.4pt}}}\ULon}

\newcommand{\be}{\begin{equation}}
\newcommand{\ee}{\end{equation}}
\newcommand{\bdm}{\begin{equation*}}
\newcommand{\edm}{\end{equation*}}
\newcommand{\bea}{\begin{eqnarray}}
\newcommand{\eea}{\end{eqnarray}}

\newcommand{\partialf}[2]
{
 \ifthenelse{\equal{#1}{}}{\frac{\partial}{\partial #2}}{\frac{\partial #1}{\partial #2}}
}

\renewcommand{\(}{\left(}
\renewcommand{\)}{\right)}

\newcommand{\Am}{\mathbf  A}

\providecommand\bnabla{\boldsymbol{\nabla}}
\providecommand\bcdot{\boldsymbol{\cdot}}

\newcounter{saveeqn}%

\def\bt{\tilde{\beta}}

\def\st{\sin{\vartheta}}

\def\Uv{\mathbf{U}}

\newcommand{\defn}{\ensuremath{\stackrel{\mathrm{def}}{=}}}
\renewcommand{\equiv}{\defn}

\providecommand\bnabla{\boldsymbol{\nabla}}
\providecommand\bcdot{\boldsymbol{\cdot}}

\newcommand{\Ret}{Re_\tau}

\renewcommand{\U}{\mathbf{U}}
\renewcommand{\u}{\mathbf{u}}

\shorttitle{Parametric mechanism and control of wall-turbulence}
\shortauthor{B. F. Farrell, P. J. Ioannou, and M.-A. Nikolaidis}

\title{Parametric mechanism maintaining Couette flow turbulence verified in DNS implies novel control strategies}

\author{Brian F. Farrell\aff{1}
\and Petros J. Ioannou\aff{1, 2}
\corresp{\email{pjioannou@phys.uoa.gr}}
\and Marios-Andreas~Nikolaidis\aff{2}}
\affiliation{\aff{1}Department of Earth and Planetary Sciences, Harvard University, Cambridge, U.S.A.
\aff{2}Department of Physics, National and Kapodistrian University of Athens, Athens, Greece}

\begin{document}

\maketitle

%
%
%
%

\date{\today}

\begin{abstract}
The no-slip boundary condition results in a velocity shear forming in fluid flow near a solid surface.  
This shear flow supports the turbulence characteristic of fluid flow near boundaries at
Reynolds numbers above $\approx 1000$  by making available to perturbations 
the kinetic energy of the externally forced
flow.  Understanding the physical mechanism underlying 
transfer of energy from the forced mean flow to the turbulent perturbation field  that is  required
to maintain turbulence   poses a fundamental question.  
Although qualitative understanding that this transfer involves nonlinear destabilization 
of the roll-streak coherent structure has been established,  identification of this instability has resisted analysis.
The reason this instability has resisted comprehensive analysis is that its analytic expression  lies in the Navier--Stokes equations (NS) 
expressed using statistical rather than state variables.  Expressing NS as a statistical state dynamics (SSD) 
at second order in a cumulant expansion suffices to allow analytical  identification of the nonlinear  
roll-streak instability underlying turbulence  in wall-bounded shear flow.   In this nonlinear instability the turbulent perturbation field is 
identified by the SSD with the Lyapunov vectors of the linear operator  governing perturbation evolution about  the 
time dependent streamwise mean flow.  In this work the implications of the
predictions of  SSD analysis that this parametric instability  underlies the dynamics of turbulence in  
Couette flow and that the perturbation structures are the associated  
Lyapunov vectors are interpreted to imply new conceptual approaches to controlling turbulence.  
It is shown that  the perturbation component of turbulence is supported on the streamwise 
mean flow, which implies optimal control should be formulated to suppress perturbations from the streamwise mean.  
It is also shown that  suppressing only the top few Lyapunov vectors on the streamwise mean vectors results 
in laminarization. These results are  verified  using DNS.

\end{abstract}


\maketitle


\section{Introduction}
Analogy with the conventional interpretation of the dynamics of isotropic homogeneous turbulence forced 
stochastically at large scale suggests that in the turbulence of wall bounded shear flows
nonlinearity leads to a cascade of  
energy from the large scales, where energy is input by  pressure gradients or boundary motions,
to small scales, where it is dissipated, and  that the turbulence field  should be essentially structureless.  
However, 
experimental studies \citep{Kline-etal-1967,Blakewell-Lumley-1967,Kim-Kline-1971,Blackwelder-Eckelmann-1979,RobinsonSK-1991,Adrian-2007}
and analysis of direct numerical simulations (DNS) \citep{Kim-etal-1987,Jimenez-Moin-1991}, have revealed distinct coherent structures in wall-turbulence which are believed to 
be essential to the process maintaining the turbulence by some form of nonlinear
regeneration cycle \citep{Kim-Kline-1971,Jimenez-1994,Hamilton-etal-1995}. This cycle involves a specific coherent structure  rather than unstructured fluctuations
and is
referred to as the self-sustaining process (SSP) ~\citep{Hamilton-etal-1995,Waleffe-1997,Jimenez-Pinelli-1999}.
These coherent structures  are streaks, that is  localized regions of increased or decreased velocity in the streamwise direction,
and quasi-cylindrical vortices with axis oriented in the streamwise direction, called rolls,  which are
collocated  with the low and high speed  streaks. The  SSP is associated with 
the low-speed streak which is produced by  lift-up   of
low speed fluid by the roll resulting in streaks which vacillate in space and time\footnote{
For the purpose of formulating the SSD used in this work we partition the flow into its streamwise mean component and perturbations.  
In this streamwise mean partition the roll-streak structure is  included in the mean flow.  
From the point of view of a partition into time or ensemble means the roll-streak would be included with the perturbation field.
However, the latter partitions do not result in expression in their associated SSD of the fundamental dynamics of wall-turbulence.}.  
Mechanistic explanations for this process posit either that 
a component of the perturbations  from the streamwise mean flow directly comprise  the roll that forces the streak
\citep{Jimenez-Pinelli-1999,Schoppa-Hussain-2000,Schoppa-Hussain-2002,Adrian-2007}  
or alternatively the roll is forced by perturbation Reynold's stresses that induce torques collocated correctly to  maintain 
the  rolls that in turn force the streaks through the lift-up process \citep{Hamilton-etal-1995}.
A number of  physical mechanisms have been invoked to address  the origin of the perturbations 
and their mechanism of action in producing this cycle. 
In one view the perturbations arise  due to  hydrodynamic instability of the streak  \citep{Waleffe-1997} 
in another they are ascribed to  growth of  highly amplifying transient perturbations in the flow \citep{Schoppa-Hussain-2002}. 
However, simply invoking such mechanisms by itself only allows
qualitative descriptions to be made for  hypothesized processes  rather 
than constituting an analytical formulation that would provide a theory based directly on the equations of 
motion with the property of making
specific testable predictions for 
observational correlates. 

Another approach invokes exact
static and periodic solutions of the Navier--Stokes equations
\citep{Waleffe-1998,Waleffe-2001,Kawahara-Kida-2001}.  These solutions have been found 
at low Reynolds numbers and shown to be at times approached 
by turbulent state trajectories (cf. \citet{Budanur-etal-2017}). 
They  provide  heuristic examples of the SSP process but these solutions
are not themselves fully turbulent states and they are not physically realizable as they  are unstable and their physical significance
to fully developed turbulence has not been established. 

The SSP was isolated recently directly from the equations of motion by showing it to be inherent to and  contained in
a highly simplified  second order closure of the Navier--Stokes equations for wall-turbulence 
\citep{Farrell-Ioannou-2012,Farrell-etal-2016-PTRSA}. 
This SSD isolates the nonlinear instability that underlies the SSP 
and the turbulence that develops under this second order
closure has been demonstrated to be 
realistic and to capture the large scale dynamics of Navier--Stokes turbulence \citep{Thomas-etal-2014,
Thomas-etal-2015, Bretheim-etal-2015,Farrell-etal-2016-VLSM,Farrell-etal-2016-PTRSA,Pausch-etal-2018}. This closure 
constitutes  a quasi-linear dynamics that greatly simplifies analysis while having
the attribute of not only allowing identification of the dynamics supporting
the large scale roll-streak  coherent structures but also analytically characterizing the  perturbation
structures responsible for maintaining the cycle.
This closure constitutes a theory for  wall-turbulence
in the  sense that it is derived from the NS, allows identification of and 
analytical solution for the dynamical mechanism underlying the turbulence 
as well as making verifiable predictions for the structure of  both the mean flow and perturbations 
as well as their specific roles in the turbulence dynamics.

In this paper  we verify  that  the perturbation structure predictions of this  second order closure
using DNS and show that these analytically characterized perturbations  are
responsible for maintaining the turbulent state and as well as that their removal leads to laminarization.  
One remarkable aspect of this identification of the perturbation structure in NS 
turbulence  
is that it is contrary
to the common assumption that
the bulk of the perturbation variance arises in association with
an energy cascade  to small scale. We show that the 
perturbation structures can be identified with the analytically fully characterized Lyapunov vectors 
sustained by the parametric growth process associated with the fluctuating mean flow.



\begin{table}
\begin{center}
\centering
\begin{tabular}{@{}*{7}{c}}
\hline
$[L_x,L_z]/h$&$N_x\times N_z\times N_y$& ${\Ret}$ &[$L_x^+$,$L_z^+]$\\
\hline
$[1.75\pi\;,\;1.2\pi]$&$33 \times 33 \times 35 $  &48.8 & $[268,184]$\\
\end{tabular}
\caption{\label{table:geometry} $[L_x,L_z]/h$ is the domain size in the streamwise, spanwise direction. 
$N_x$, $N_z$ are the number of Fourier components after dealiasing with the $1/3$ rule 
and $N_y$ is the number of equally spaced points in the wall-normal direction. $\Ret=u_\tau h/\nu$ 
 is the Reynolds number based on the friction velocity $u_\tau = \nu \left . du/dy \right |_{y=h}$,
   and $[L_x^+$,$L_z^+]$ is the channel size in wall units $\nu/u_\tau$.}
  \end{center} 
\end{table}

\section{Formulation}
We illustrate these results using DNS for the case of Couette flow turbulence at
Reynolds number $R=600$   ($R=U_w h/ \nu$,  
where $\pm U_w$ is the velocity at the channel walls at $y=\pm h$
and $\nu$ is the coefficient of kinematic viscosity).  
The laminar Couette flow is in the  $x$-direction (the streamwise direction) and is given by
$\u = (U_w y/h,0,0)$, $y$, the second component,  is the cross-stream direction, and $z$, the third component, is the 
spanwise direction. 
The details of the direct numerical simulation are given in Table \ref{table:geometry}.

We formulate 
the second order SSD equations  by decomposing
the flow field into its streamwise mean component, denoted by $\langle \cdot \rangle_x$ or alternatively by capital letters, and the deviations from 
the streamwise mean,
referred to as perturbation components and denoted with a dash,  or equivalently  into the $k_x=0$ and the $k_x \ne 0$ components
of the  Fourier decomposition of the flow field, where $k_x$ is the
streamwise wavenumber:
\begin{equation}
\u = \Uv(y,z,t) + \u'(x,y,z,t)~,~~~\Uv(y,z,t) \equiv \langle \u \rangle_x~.
\label{eq:mean}
\end{equation}

The  Navier--Stokes equations for incompressible flow expressed using this mean and perturbation partition are:
\begin{subequations}
\label{eq:NS}\begin{gather}
\partial_t\U + {\U  \bcdot \bnabla  \U }   + \bnabla  P/\rho -  \nu \Delta \U  = {- \langle\u ' \bcdot \bnabla  \u '\rangle_x}\ ,
\label{eq:NSm}\\
 \partial_t\u '+  { \U  \bcdot \bnabla  \u ' +
\u ' \bcdot \bnabla  \U } + \bnabla   p' /\rho-  \nu \Delta  \u ' = \underbrace{- \( \u ' \bcdot \bnabla  \u ' - \langle\u ' \bcdot \bnabla  \u '\rangle_x \,\)}_{N} ~,
 \label{eq:NSp}\\
 \bnabla  \bcdot \U  = 0\ ,\ \ \ \bnabla  \bcdot \u ' = 0\ ,
 \label{eq:NSdiv0}
\end{gather}\label{eq:NSE0}\end{subequations}
where $P$ is the pressure and $\rho$ the constant density.
We study a turbulent Couette flow confined  in a doubly periodic channel in $x$ and $z$ satisfying no-slip boundary conditions in the cross-stream direction: $\U (\pm h,z,t)= (\pm U_w,0,0)$,
$\u' (x,\pm h,z,t)= (0,0,0)$.
The mean velocity has three components $\U(y,z,t)=(U,V,W)$, with the cross-stream velocity $V$ and spanwise velocity 
$W$
expressible by a streamfunction as $V=-\partial_z \Psi$, $W= \partial_y \Psi$.

The SSD we use is closed at second order  by simply setting the third cumulant to zero which is equivalent 
to ignoring the perturbation-perturbation nonlinearity, $N$, in \eqref{eq:NSp}  when formulating the SSD \citep{Herring-1963,Farrell-Ioannou-2003-structural}. 
If the same term is ignored in the partition of the NS into mean and perturbations this produces 
what is referred to as the restricted nonlinear  (RNL) system  \citep{Thomas-etal-2014,Farrell-etal-2016-PTRSA}:
\begin{subequations}
\label{eq:RNL}
\begin{gather}
\partial_t\U + {\U  \bcdot \bnabla  \U }   + \bnabla  P/ \rho -  \nu \Delta \U  = {- \langle\u ' \bcdot \bnabla  \u '\rangle_x}\ ,
\label{eq:RNLm}\\
 \partial_t\u '+  { \U  \bcdot \bnabla  \u ' +
\u ' \bcdot \bnabla  \U } + \bnabla   p' / \rho  -  \nu \Delta  \u ' = 0 \, \label{eq:RNLp}\\
 \bnabla  \bcdot \U  = 0\ ,\ \ \ \bnabla  \bcdot \u ' = 0\ .
 \label{eq:RNLdiv0}
\end{gather}\end{subequations}
This RNL system has the same quasi-linear structure as the second order SSD and
can be regarded as an approximation to the second-order SSD 
in which one ensemble member is used to obtain 
the second cumulant and it has a number of interesting properties \citep{Farrell-Ioannou-2017-sync}. 
One of these is that this system supports realistic turbulence 
despite the absence of nonlinearity $N$, 
which provides a constructive proof that this  explicit perturbation 
nonlinearity is not responsible for maintaining turbulence.  
A second remarkable implication is that 
analytical identification of 
the perturbation structure and dynamics  follows directly from analysis of \eqref{eq:RNLp}.
Consider a self-sustaining turbulent solution of \eqref{eq:RNL} with mean flow  $\U(y,z,t)$.
The associated perturbation field consistently satisfies \eqref{eq:RNLp} 
with this $\U(y,z,t)$, which means that  the perturbations evolve according to
the time-dependent linear operator 
\eqref{eq:RNLp}, or symbolically: $\partial_t \u' =  \Am ( \U) \u'$, with $\Am$ this time 
dependent linear operator. This allows complete identification of the perturbation field 
with the Lyapunov vectors of $\Am (\U)$.  
Moreover, because the  turbulence supported by \eqref{eq:RNLp}
is bounded and nonzero it follows that
$\u'$ must lie in the restricted subspace spanned by the Lyapunov vectors 
of~$\Am (\U)$   with zero Lyapunov exponent because if the Lyapunov 
exponent is positive the associated vector would become unbounded 
and if negative it  would vanish. Integration of the RNL 
system \eqref{eq:RNL} reveals that even at moderately high Reynolds numbers 
this subspace is supported by a small set of streamwise 
harmonics~\footnote{Because $\U$ 
is independent of $x$ each Lyapunov vector is supported by a single streamwise wavenumber. For a discussion of the
streamwise harmonic support of the Lyapunov vectors  cf. \citep{Thomas-etal-2015,Farrell-etal-2016-VLSM}.},
and consequently RNL turbulence is supported solely by  these few 
harmonics, which provides a constructive identification of the active subspace underlying this turbulence.  
For example, the RNL turbulence of Couette flow 
at $R=600$ in the present channel is supported by a single
Lyapunov vector with the gravest non-zero streamwise wavenumber 
$k_x = 2 \pi / L_x$ \citep{Farrell-Ioannou-2017-sync}.  
Consistently, in a realization of RNL turbulence only 
this perturbation component survives.
In summary, we have obtained
a full analytic characterization of the perturbation field that sustains RNL
turbulence: it is the subspace of the Lyapunov vectors
of $\Am (\U)$  with zero  Lyapunov exponent.
It is important to note that each of the ingredients of this turbulence are characterized:
the coherent structures are the streamwise mean streaks, defined as the  streamwise 
mean velocity that obtains after removal of its spanwise average: 
$U_s \equiv U(y,z,t) - \langle  U(y,z,t) \rangle_z$, the rolls with
streamfunction $\Psi$ collocated with the streaks 
and finally the Lyapunov vectors of operator $\Am (\U)$ with zero exponent,
which are analogous to  neutral modes of a time independent linear operator.  In RNL removal 
of the subspace of the Lyapunov vectors of~$\Am (\U)$ with  zero 
Lyapunov exponent leads immediately to laminarization \footnote{ 
Note that these
are not the Lyapunov vectors of the trajectory of the 
full non-linear system governing the mean and the perturbations}. We now show that removing only a 
few of the least stable LVs of the streamwise mean flow  in a DNS suffice to laminarize the 
turbulence in our channel at $R=600$.  

\begin{figure}
\centering
\includegraphics[width=0.7\columnwidth]{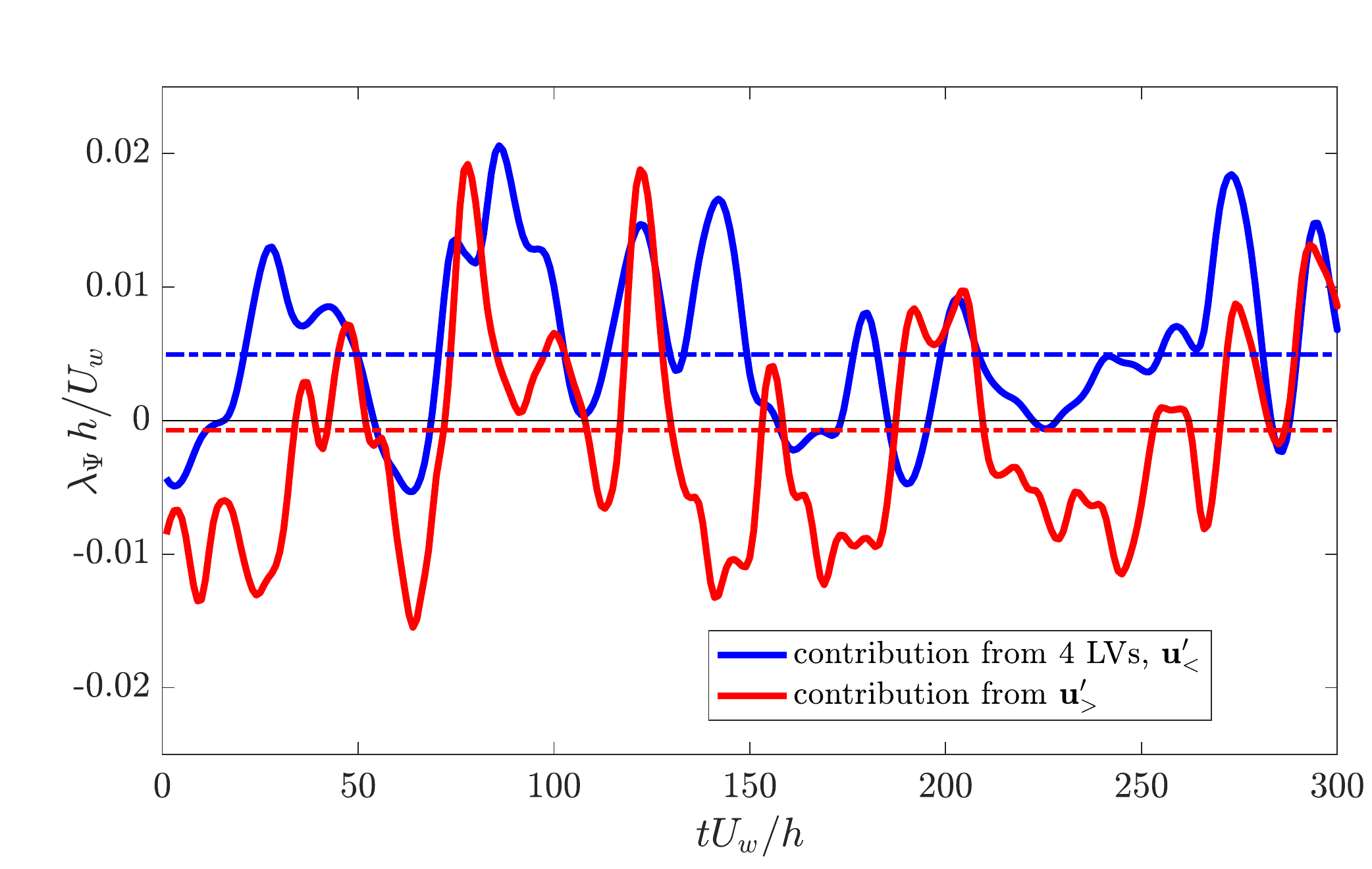}
\caption{Time evolution of the contribution of the torques arising from the Reynolds stresses produced by  $\u'_<$  to  maintenance of $\Psi^2$, which largely consists of the rolls (blue). 
The torques from $\u'_>$ (red) make no net contribution to the rolls in this measure.
Their time mean  is indicated with  dashed lines.
This figure identifies the perturbation subspace responsible for maintaining the roll against dissipation to be the subspace spanned by the four least stable  LVs.} 
\label{fig:Psi}
\end{figure}


\section{Results}
The six least stable LVs of the linear operator~$\Am$ about of the mean flow $\U(y,z,t)$ of the turbulent state  
at $R=600$  have Lyapunov exponents:  
$$ (0.02, 0.007, -0.0002, -0.0056, -0.013, -0.017)\,U_w/h\ .$$ The Lyapunov vectors and exponents are calculated by evolving in parallel with the DNS simulation Eq. \eqref{eq:RNLp} with the $\U(y,z,t)$ obtained from the DNS. Using the standard power method and successive orthogonalizations
using the energy inner-product we obtain the Lyapunov vectors of the $\U(y,z,t)$ and their characteristic exponents.

As in the RNL turbulence all of these least stable LVs are found to be supported by the 
gravest streamwise wavenumber permitted in the channel
$k_x=2 \pi/L_x$ \footnote{Although in RNL the Lyapunov exponents 
are necessarily $\le 0$ when \eqref{eq:RNLp} is used with the $\U$ of the RNL, inclusion of the 
$N$ term in \eqref{eq:RNLp}  produces a small component of energy loss  by these vectors so that consistently
the top Lyapunov exponent exceeds zero by this amount \citep{Nikolaidis-etal-Madrid-2018}.
}.
The perturbation structure in a DNS can be projected on the basis of the
orthogonalized LVs. Doing so we find that the perturbations in the DNS 
have significant projection
on the first LV ($11 \%$ on average) and about $20 \%$ on average on the subspace spanned by the 
four least stable LVs.  These least stable  Lyapunov vectors also 
dominate the others in the rate of energy extraction from the streamwise flow $U(y,z,t)$,
so we anticipate that removal of these vectors should produce a severe restriction of the flow of energy from
the mean flow to the perturbations.  More remarkable for our study than  dominance 
of the energetics of the perturbations by these four least  stable LVs 
is that they  account fully for the forcing of the roll and 
therefore the SSP cycle. In order to assess the contribution of the Lyapunov vectors to the roll forcing
consider the equation for the streamwise component  $\Omega_x = \Delta_h \Psi$ with $\Delta_h \equiv \partial_y^2+\partial_z^2$,
of the mean  vorticity equation:
\begin{align}
\partial_t \Omega_x & = \underbrace{ -\left ( V \partial_y + W\partial_z \right ) \Omega_x}_A+ \underbrace{ \nu \Delta_h \Omega_x}_{D} \underbrace{- \left [ (\partial^2_{y}-\partial^2_{z} )\langle vw \rangle_x  + \partial_{yz} \left ( \langle w^2 \rangle_x - \langle v^2 \rangle_x  \right ) \right ] }_{G_{\Omega_x}} ~,
\label{eq:MPSI}
\end{align}
Terms $A$ and $D$  represent advection and dissipation of $\Omega_x$ in the $(y-z)$ plane
and if it were not acted upon by the streamwise mean torque from the perturbation Reynolds stresses,  $G_{\Omega_x}$,  the roll
would decay.
The contribution of perturbation Reynolds stresses  to the rate of change of the normalized
 streamwise square vorticity 
can be measured  by $\lambda_{\Omega_x} = {\int_V {\Omega_x} G_{\Omega_x} dV}/({2 \int_V  \Omega_x^2 dV})$,
and similarly, if more emphasis is to be given to the large scales,
we could use as a measure  the contribution of the perturbation Reynolds stresses
to the maintenance of  the square of	 the streamfunction.
This normalized measure of contribution to $\Psi^2$ is
$\lambda_{\Psi} = {\int_V {\Psi} G_{\Psi} dV}/({2 \int_V  \Psi^2 dV})$,
where $G_\Psi \equiv \Delta_h^{-1} G_{\Omega_x}$ and $\Delta_h^{-1}$ is the inverse 
 cross-stream/spanwise Laplacian.
In this work we decompose the perturbation field $\u'$ into its 
component, $\u_{<}'$, projected on the subspace spanned by the 4 least damped energy orthonormal LVs,
denoted $\u_i'$, $i=1,2,3,4$ and the projection on the complement $\u_>'$:
\begin{equation}
\u_{<}' \equiv \sum_{i=1}^4 (\u'\cdot \u_i' ) \u_i'~,~~~\u_{>}' \equiv \u'-\u_{<}'~,
\end{equation} 
and estimate  $G_\Psi$ produced by $\u_<'$ and $\u_>'$. 
The contribution of these subspaces to  $\lambda_\Psi$ is shown  in Fig. \ref{fig:Psi}.
It can be seen that the first four least stable LVs contribute $100 \% $ on average 
to  the roll maintenance\footnote{The first LV contributes to $\lambda_\Psi$ 
on average $60\%$, while inclusion of the second LV adds another $26 \%$. 
The corresponding contribution to
$\lambda_{\Omega_x}$ by $\u_<'$ is $20 \%$ consistent with more emphasis 
being placed on small scale vorticity by the square vorticity measure.}.
\begin{figure}
\centering
\includegraphics[width=0.7\columnwidth]{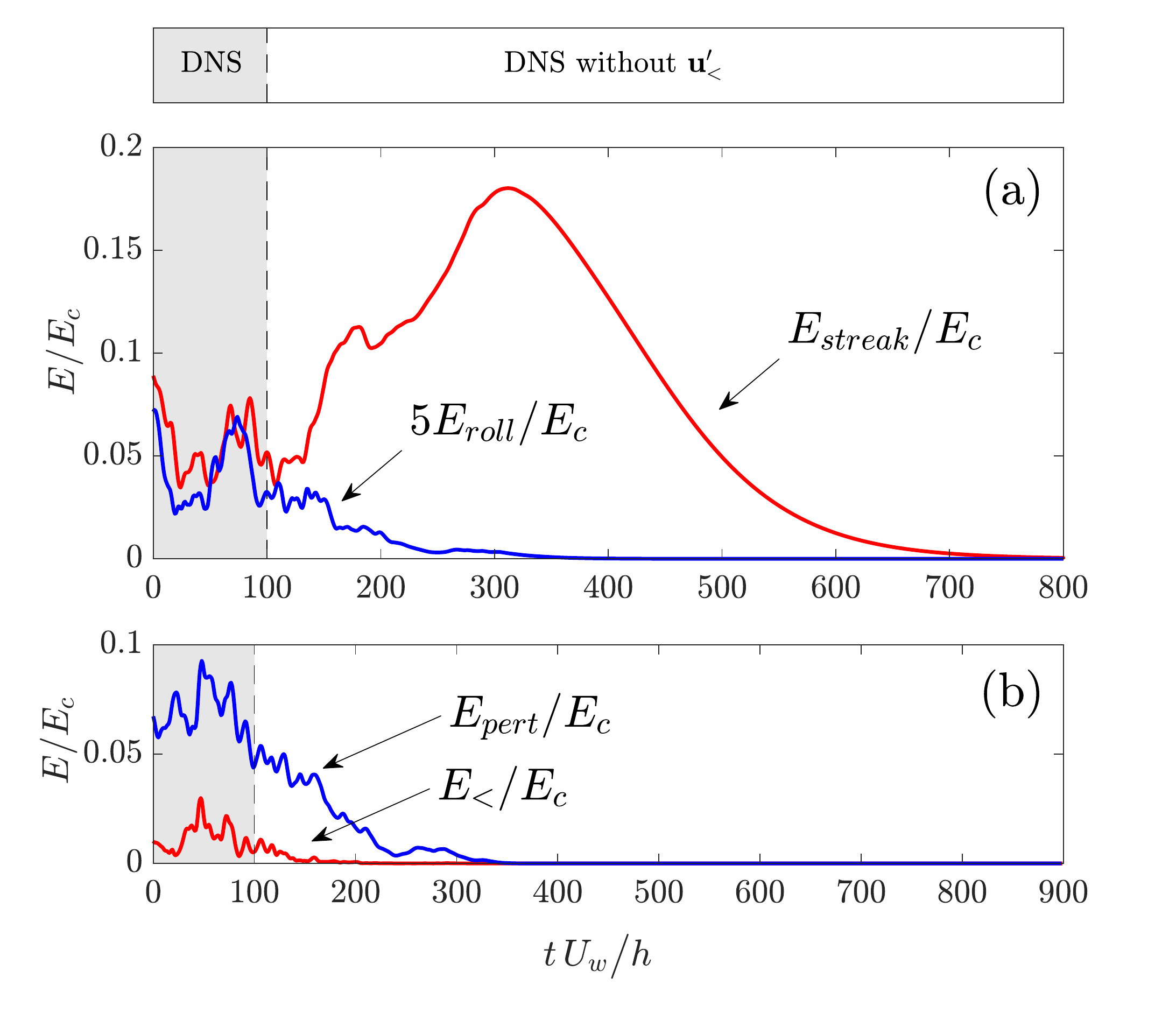}
\caption{(a): Evolution of the streak energy $E_{streak}=\int_V U_s^2 dV/2$, of the roll energy   $ E_{roll} = \int_V (V^2+W^2) dV/2$ (multiplied by 5). At $t=100h/U_w$  the component, $\u_<'$, that 
 projects on the subspace spanned by the four least stable LVs is gradually
 removed. (b): Evolution of 
 the perturbation energy $E_{pert}=\int_V |\u'|^2 dV / 2$  and the energy of the component
  that lies in the subspace of the four least stable LVs $E_<=\int_V |\u_<'|^2  dV/2 $. 
  All energies are normalized by the energy of the laminar Couette flow, $E_c$.} 
\label{fig:lamin}
\end{figure}

\begin{figure}
\centering
\includegraphics[width=0.7\columnwidth]{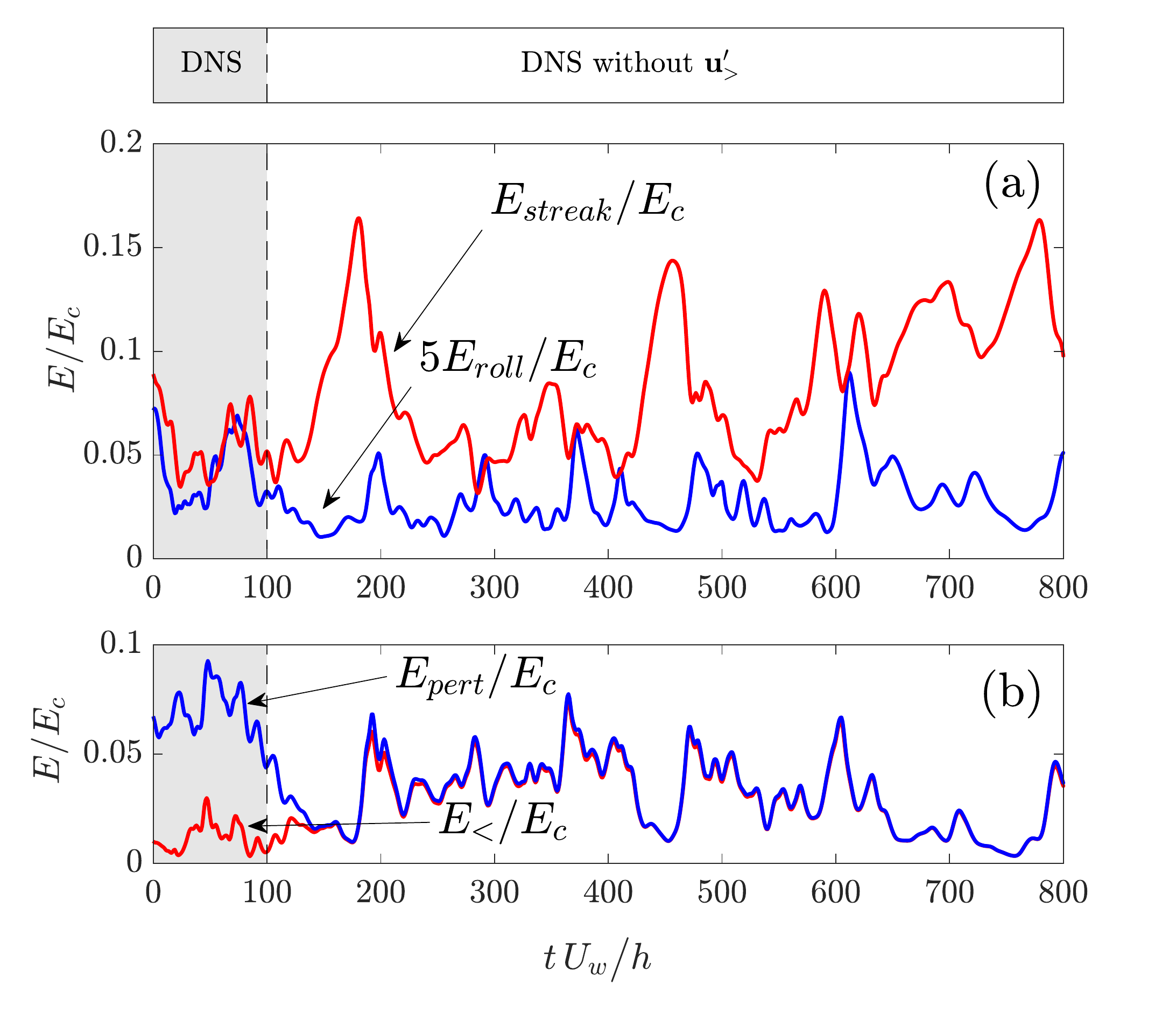}
\caption{(a): Same as Fig. \ref{fig:lamin} but now the complement $\u_>'$ is removed with the same protocol. 
The turbulence sustains in DNS with the full perturbation field
converging to a single Lyapunov vector of the fluctuating mean flow.}  
\label{fig:nolamin}
\end{figure}

This identification of a small subset of the least stable  LVs
as the perturbation structures that support  the SSP  anticipates
laminarization of the turbulence in the DNS upon removal of this subspace.   A long integration of the 
DNS of  Couette turbulence at $R=600$ has been used to obtain the converged structures of the four least stable LVs. At a specified time
the perturbation field, $\u'$, of the DNS 
is projected on the  $\u_<'$. 
We remove this component of the perturbation field at an increasing rate $f(t)$
so that the perturbation field at each time step becomes $ \u'- f(t) \delta t \u_{<}'$, where $\delta t$ is the integration 
time step.  In the example  
shown in Fig. \ref{fig:lamin} this process of gradual removal of the first four Lyapunov vectors starts  at $t=100 h/U_w$, so that $f(t)=0$ for $t <100h/U_w$ and the rate  $f(t)$ increases linearly from $0$ to $1$
at $t=300 h/ U_w$,  with $f=1$ after this time. With the gradual removal of this subspace
 the entire perturbation field as well as the rolls decay  leading to laminarization.
 The streak is seen to increase in magnitude before eventually decaying, as is typical of laminarization events
 due to the energy extraction from the streak having been suppressed by the loss of the 
 perturbations while the roll  remains relatively more effective at continuing the lift-up process forcing the streak.
  An alternative experiment was performed in which the complement $\u_>'$ 
  of the four least damped Lyapunov vectors  
  was removed with the same protocol
 and the turbulence was shown to sustain in DNS with the perturbation structure 
 thus constrained to lie in the subspace
 of the four least damped LVs. The corresponding evolution of the energies of this 
 experiment are shown in Fig. \ref{fig:nolamin}.
 The turbulence that results approaches the corresponding
 RNL turbulence, differing only 
 in that the perturbation-perturbation nonlinearity introduces an additional sink for the perturbation energy. 
 The perturbation field collapses to the single top LV of the fluctuating mean flow, as is the case
 for RNL turbulence in the same channel and Reynolds number \citep{Farrell-Ioannou-2017-sync}.
  The turbulence that results supports rolls of approximately the same 
 magnitude while the streaks become stronger
 as the streak is embedded in an environment of reduced eddy viscosity with the removal of the 
 higher LVs. In the supplemental material we have included 
 video showing the turbulence of Fig. \ref{fig:nolamin} and
 the laminarization of Fig. \ref{fig:lamin}.

%
%

\section{Conclusions} 

In this work we have verified in a DNS of turbulent Couette flow  
a number of 
the predictions of a second order  SSD comprising the streamwise 
mean  flow and the second cumulant  of the perturbation field closed by neglecting the third 
cumulant.  These predictions  include  identification of the  
mean and perturbation structures as well as  
of the physical mechanism  supporting the mean flow and the perturbations.  
The mechanistic component of the SSP that is responsible for supporting 
the perturbation field and collocating it with the streak so as to 
maintain the fluctuating streak SSP is identified with the parametric instability of the fluctuating 
streak and while the first four  Lyapunov vectors of this 
instability have been verified to
account for 20 \% of the energy of the perturbation 
field  it is more significant for our purposes that they account for all of the torque 
supporting the roll component of the roll-streak structure underlying the SSP 
maintaining the turbulence.  Consistently, 
removal of  this small subset of structures is verified to laminarize the turbulence in the DNS.

It is a common assumption that the structure and maintenance of the perturbation field at 
scales smaller than the integral can be ascribed to a nonlinear cascade 
and therefore that these scales can be characterized solely by their spectrum.  
In this work we have shown that in Couette flow turbulence this is not the case and that 
these perturbations are primarily maintained by parametric interaction with 
the mean flow   and that their structure rather than being random can be identified with
the Lyapunov vectors associated with this parametric growth process.

These results imply that optimal control strategies based on linearization 
about the time-dependent
streamwise mean flow, which is the first cumulant of the RNL statistical state dynamics, should present substantial advantage
over  previous optimal control strategies that were based on the instantaneous streamwise and spanwise mean flow
\citep{Bewley-Liu-98,Hogberg-etal-03a,Hogberg-etal-03b,Hogberg-etal-2003, Kim-Bewley-2007}.
This implication is strengthened by simulations confirming that turbulence is not 
supported when the fluctuating streaks that are present in the streamwise 
mean flow are suppressed \citep{Jimenez-Pinelli-1999}. 
Perhaps most remarkable is the result that a very small subspace of 
perturbations are responsible for supporting the roll circulation 
required for maintaining the turbulence.  This subspace is much 
smaller than that maintaining the perturbation variance 
demonstrating  that suppression of a small subspace of the entire 
perturbation field is sufficient to laminarize the turbulence. 


\vspace{1em}

This work was done  during the 2018 Center for Turbulence Research Summer Program
with financial support from Stanford University and NASA Ames Research Center. We would like
to thank Professor~Parviz~Moin, Professor Javier Jim\'enez, Dr.~Adri\'an Lozano-Dur\'an,  Dr.~Michael Karp 
and Dr. Navid Constantinou for their useful comments and discussions. Marios-Andreas Nikolaidis gratefully acknowledges the support of the Hellenic Foundation for Research and Innovation (HFRI) and the General Secretariat for Research and Technology (GSRT). Brian F. Farrell  was partially supported by NSF AGS-1640989.

\end{document}